\documentclass[aps,preprint,flotfix]{revtex4-2}
\usepackage{latexsym}
\usepackage{graphics}
\usepackage{graphicx,amssymb,epsfig,epsf,amsmath}
\usepackage{physics}
\usepackage[colorlinks]{hyperref}

\def\be{\begin{equation}}
\def\lan{\left\langle}
\def\ran{\right\rangle}
\def\ee{\end{equation}}
\def\barr{\begin{array}}
	\def\earr{\end{array}}

\def\l{\left}
\def\r{\right}
\def\dis{\displaystyle}
\def\ed{\end{document}}

\def\kp{{\kappa}}

\begin{document}
	
	\title{Structure of wavefunction for interacting bosons in mean-field with random $k$-body interactions}
	
	\author{Priyanka Rao}
	\author{N. D. Chavda\footnote{ndchavda-apphy@msubaroda.ac.in}}
	
	\affiliation{Department of Applied Physics, Faculty of Technology and
		Engineering, The Maharaja Sayajirao University of Baroda, Vadodara-390001, India}
	
\begin{abstract}
	Wavefunction structure is analyzed for dense interacting many-boson systems using Hamiltonian $H$, which is a sum of one-body $h(1)$ and an embedded GOE of $k$-body interaction $V(k)$ with strength $\lambda$. In the first analysis, a complete analytical description of the variance of the strength function as a function of $\lambda$ and $k$ is derived and the marker $\lambda_t$ defining thermalization region is obtained. In the strong coupling limit ($\lambda > \lambda_t$), the conditional $q$-normal density describes Gaussian to semi-circle transition in strength functions as body rank $k$ of the interaction increases. In the second analysis, this interpolating form of the strength function is utilized to describe the fidelity decay after $k$-body interaction quench and also to obtain the smooth form for the number of principal components, a measure of chaos in finite interacting many-particle systems. The smooth form very well describes embedded ensemble results for all $k$ values.
\end{abstract}
	

\maketitle
\date{}

\section{Introduction}
\label{sec:1}
It is now well established that Random Matrix Theory, due to it's universality \cite{BGS-84}, successfully describes the  spectral as well as wavefunction properties of isolated finite many-particle quantum systems \cite{kota-book}. The spectral statistics deals only with the energy
eigenvalues while 
the statistical properties related to the structure of the wavefunctions can reveal different layers of chaos and hence give profound understanding of various problems in the field of quantum many-body chaos and thermalization, in isolated finite interacting particle systems such as atomic nuclei, atoms, mesoscopic systems (quantum dots, small metallic grains), interacting spin systems modeling quantum computing core, ultra-
cold atoms and quantum black holes with SYK model and so on \cite{kota-book,SenRMP,Rigol,BISZ2016,KC-18, Cotler-2017,verbaarschot_QCD,kc-18b}. To
analyze the wavefunction properties, it is very crucial to examine the so-called strength functions (also known as local density of states) in detail, as they give information about how a particular basis state spreads onto the eigenstates. The chaos measures like number of principal components (NPC), information entropy, fidelity decay etc. can also be determined by examining the general features of the strength functions \cite{kota-book}.

The statistical properties of isolated finite many-particle quantum systems investigated by employing random
matrix ensembles are generally referred as Gaussian ensembles (and in particular the Gaussian orthogonal ensemble (GOE)) for $m$-particle system. They involve interaction up to $m$-body in character and are dominated by the $m$-body interactions. However, constituents of isolated quantum systems interact via few-body interactions. 
Hence the concept of embedded
ensemble (EE) of $k$-body interaction, in particular EGOE($k$) (GOE version of EE($k$)) was introduced by French and co-workers \cite{FW70,BF71}. These models for the
particles in a mean-field and interacting via two-body interactions ($k=2$) and their various extended versions form good
models for understanding various aspects of chaos in interacting
particle systems \cite{kota-book} and they are investigated in detail both for fermion systems (called EGOE(1+2)) \cite{vkb2001,Brody,PW-07,FI1,FI2,MF} as well as boson systems (called BEGOE(1+2) with 'B' for bosons) \cite{Pa-00,Ag-01,Ag-02,Ch-03,Ch-04,Ch-PLA}. Here, with $m$ particles distributed in $N$ single particle (sp) states, two limiting situations exist, one is the dilute limit (defined as $m\rightarrow\infty$, $N \rightarrow \infty$ and $m/N \rightarrow 0$) and another is the dense limit (defined by $m\rightarrow\infty$, $N \rightarrow \infty$ and $m/N \rightarrow \infty$). In the dilute limit, one can expect similar behavior for both fermion and boson systems while the dense limit is feasible only for boson systems and therefore the focus was on the dense limit in BEGOE investigations \cite{Pa-00,Ag-01,Ag-02,Ch-03,Ch-04,Ch-PLA,CK2017}. For EGOE(1+2) in dilute limit and for BEGOE(1+2) in dense limit, as a function of the two-body interaction
strength $\lambda$ (measured in units of the average spacing between the
one-body mean-field sp  levels), exhibits three transition or chaos
markers $(\lambda_C,\lambda_F,\lambda_t)$: (a)~as the two-body interaction is
turned on, level fluctuations exhibit a transition from Poisson to GOE at
$\lambda=\lambda_C$; (b) with further increase in $\lambda$, the strength
functions make a transition from
Breit-Wigner (BW) form to Gaussian form at $\lambda=\lambda_F > \lambda_C$; and (c)
beyond $\lambda=\lambda_F$, there is a region of thermalization around
$\lambda=\lambda_t$ where the basis dependent thermodynamic quantities like entropy behave alike. It is important to note that
the transitions mentioned above are inferred from large number of numerical calculations and they are well verified
to be valid in the bulk part of the
spectrum. For further details see~\cite{kota-book} and references there in. 

Going beyond two-body interaction, it is seen that the higher body interactions i.e. $k>2$ play an important role in strongly interacting quantum systems \cite{Blatt,Hammer}, nuclear physics \cite{Launey}, quantum black holes \cite{Cotler-2017,Garcia-2018} and wormholes \cite{Garcia-2019} with SYK model and also in quantum transport in disordered networks connected by many-body interactions \cite{centro-1,centro-2,Ortega2018}. Therefore, it is necessary to extend the analysis of EE to higher $k$-body interactions in order to understand these problems. From the previous studies, it is known that with EGOE($k$) or (BEGOE($k$)), the eigenvalue density for a system of $m$ fermions/bosons in $N$ sp states changes from Gaussian form to semi-circle as $k$ changes from 2 to $m$ \cite{kota-book,KC-18,Brody,Manan-Ko}. Very recently, $q$-Hermite polynomials have been employed to study spectral densities of the so-called SYK model \cite{GaVerb,Verb-qhp} and quantum spin glasses \cite{spin-g}, along with studying the strength functions and fidelity decay (also known as survival or return probability) in EE, both for fermion as well as boson systems \cite{Manan-Ko}. The smooth form of eigenvalue density can be given by the so-called $q$-normal distribution $f_{qN}$ and formulas for parameter $q$ in terms of $m$, $N$ and $k$ are derived for fermionic and bosonic EE($k$) in \cite{Manan-Ko} which explain the Gaussian to semi-circle transition in spectral densities, strength functions and fidelity decay in many-body quantum systems as a function of rank $k$ of interactions. Recently, the lower-order bivariate reduced moments of the transition strengths are examined for the action of a transition operator on the eigenstates generated by EGOE($k$) and it is shown that the ensemble averaged distribution of transition strengths follows a bivariate $q$-normal distribution $f_{biv-qN}$ and a formula for NPC in the transition strengths from a state is obtained \cite{KM2020}. Very recently, analytical formulas for the lowest four moments of the strength functions for fermion systems modeled by EGOE(1+$k$) are derived and it is shown that the conditional $q$-normal density $f_{CqN}$ can be used to represent strength functions in the strong coupling limit \cite{KM2020c}. One can expect similar behavior for isolated finite interacting boson systems with $k$-body interactions in the dense limit. The purpose of the present letter is firstly to demonstrate that in strong coupling domain (in the thermalization region), the strength functions indeed can be represented by the conditional $q$-normal distribution $f_{CqN}$ in the dense interacting boson systems interacting via $k$-body interaction. Secondly, using $f_{CqN}$ form and parameters that enter in this form, fidelity decay is described in BEGOE(1+$k$) and an analytical formula for NPC is derived. 

The Letter is organized as follows. We briefly introduce BEGOE(1+$k$) and $q$-Hermite polynomials along with their generating function and conditional $q$-normal distribution in Section~\ref{sec:2}. The numerical results of the variation of parameter $q$ as a function of $k$-body interaction strength $\lambda$ in  BEGOE(1+$k$) are presented in Section~\ref{sec:3}. Also the formula of $q$ for BEGOE($k$) is given for the sake of completeness, even though it is clearly given in \cite{KC-18,Manan-Ko}. Further, a complete analytical description of the variance of the strength function, in terms of the correlation coefficient $\zeta$, for BEGOE(1+$k$) is given and ($m$,$N$,$k$) dependence of marker $\lambda_t$ is derived.  In Section \ref{sec:4}, the results for the variation of strength function, in the strong coupling domain ($\lambda >>\lambda_t$), are presented as a function of body rank $k$ and ensemble averaged results are compared with smooth forms given by $f_{CqN}$. In Section~\ref{sec:5} the interpolating form $f_{CqN}$ for the strength function is utilized to describe the fidelity decay after random $k$-body interaction quench in BEGOE(1+$k$) in the thermalization region. Further, two parameter ($\zeta$ and $q$) analytical formula for NPC is derived as a function of energy for $k$-body interaction and tested with numerical embedded ensemble results in Section~\ref{sec:6}. Finally, the concluding remarks are given in section~\ref{sec:7}.

\section{Preliminaries}
\label{sec:2}

\subsection{Embedded bosonic ensembles - BEGOE(1+$k$)}

\label{sec:2a}

Consider $m$ spinless bosons distributed in $N$ degenerate sp states interacting via $k$-body ($1 \leq k \leq m$) interactions. Distributing these $m$ bosons in all possible ways in $N$ sp states generates many-particle basis of dimension $d={N+m-1 \choose {m}}$. The $k$-body random Hamiltonian $V(k)$ is defined as,
\be
V(k) = \dis\sum_{k_a,k_b} V_{k_a,k_b} B^\dagger(k_a) B(k_b)\;.
\label{eq.ent1}
\ee
Here, operators $B^\dagger(k_a)$ and $B(k_b)$ are $k$-boson creation and annihilation operators. They obey the boson commutation relations. $V_{k_a,k_b}$ are the symmetrized matrix elements of $V(k)$  in the $k$-particle space with the matrix dimension being $d_k={N+k-1 \choose k}$. They are chosen to be randomly distributed independent Gaussian variables with zero mean and unit variance, in other words, $k$-body Hamiltonian is chosen to be a GOE. BEGOE($k$) is generated by action of $V(k)$ on the many-particle basis states. Due to $k$-body nature of interactions, there will be zero matrix elements in the many-particle Hamiltonian matrix, unlike a GOE. By construction, we have a GOE for the case $k = m$. For further details about these ensembles, their extensions and applications, see \cite{kota-book,BRW,Ma-Th} and references therein.

In realistic systems, bosons also experience mean-field generated by presence of other bosons in the system and hence, it is more appropriate to model these systems by BEGOE($1+k$) defined by,
\be
H= h(1) + \lambda V(k)
\label{eq.ent2}
\ee
Here, the one-body operator $h(1)=\sum_{i=1}^N \epsilon_i n_i$ is described by fixed sp energies $\epsilon_i$; $n_i$ is the number operator for the $i$th sp state. The parameter $\lambda$ represents the strength of the $k$-body interaction and it is measured in units of the average mean spacing of the sp energies defining $h(1)$. In this analysis, we have employed fixed sp energies $\epsilon_i = i + 1/i$ in defining the mean-field Hamiltonian $h(1)$. As the dense limit is more interesting for bosons, for numerical study, we have chosen $N = 5$, $m = 10$ with space dimensionality of $d=1001$ and varied $k$ from 2 to $m$. It is now known that in nuclear reactions and strongly interacting quantum systems $k=2,3,4$ are of physical importance\cite{Cotler-2017,Blatt,Hammer}. However for the sake of completeness, to study the generic features of embedded ensembles and the possibility of higher $k$ becoming prominent, we address $k=2$ to $m$.

\subsection{$q$-Hermite polynomials and conditional $q$-normal distribution}

\label{sec:2b}

The $q$-Hermite polynomials were first introduced by L. J. Rogers in Mathematics. Consider $q$ numbers $[n]_q$ defined as $\l[n\r]_q = (1-q)^{-1}(1-q^n)$. Then, $[n]_{q \rightarrow 1}=n$, and $[n]_q!=\Pi^{n}_{j=1} [j]_q$ with $[0]_q!=1$. Now, $q$-Hermite polynomials $H_n(x|q)$  are defined by the recursion relation \cite{ISV-87},
\be
x\,H_n(x|q) = H_{n+1}(x|q) + \l[n\r]_q\,H_{n-1}(x|q)
 \ee
with $H_0(x|q)=1$ and $H_{-1}(x|q)=0$. Note that for $q=1$, the $q$-Hermite polynomials reduce to normal Hermite polynomials (related to Gaussian) and for $q=0$ they will reduce to Chebyshev polynomials (related to semi-circle). Importantly, $q$-Hermite polynomials are orthogonal within the limits $\pm 2/\sqrt{1-q}$, with the $q$-normal distribution $f_{qN}(x|q)$ as the weight function defined by \cite{KM2020},
\be
f_{qN}(x|q)  =  \dis\frac{\sqrt{1-q}}{2 \pi \sqrt{4-(1-q)x^2}} \dis\prod_{i = 0}^{\infty} (1-q^{i+1}) [(1+q^i)^2-(1-q)q^i x^2].
\label{eq.ent7}
\ee
Here, $-2/\sqrt{1-q} \leq x \leq 2/\sqrt{1-q}$ and $q \in [0,1]$. Note that $\int_{s(q)} f_{qN}(x|q)\;dx=1$ over the range $s(q)=(-2/\sqrt{1-q}, 2/\sqrt{1-q})$. It is seen that in the limit $q \rightarrow 1$, $f_{qN}(x|q)$ will take Gaussian form and in the limit $q=0$ semi-circle form. Now the bivariate $q$-normal distribution $f_{biv-qN}(x, y|\zeta, q)$ is defined as follows \cite{KM2020,SZAB},

\begin{equation}
\barr{rcl}
f_{biv-qN}(x, y|\zeta, q) &=& f_{qN}(x|q) f_{CqN}(y|x;\zeta,q)\\\\
 &=&f_{qN}(y|q) f_{CqN}(x|y;\zeta,q)
\earr
\label{eq:bivq}
\end{equation}
where $\zeta$ is the bivariate correlation coefficient and the conditional $q$-normal densities, $f_{CqN}$ can be given as,
\begin{equation}
\barr{rcl}
 f_{CqN}(x|y;\zeta,q) &=& f_{qN}(x|q) \; \dis\prod_{i=0}^{\infty} \frac{(1-\zeta^2 q^i)}{h(x,y|\zeta,q)};\\\\

 f_{CqN}(y|x;\zeta,q) &=& f_{qN}(y|q) \;\dis\prod_{i=0}^{\infty} \frac{(1-\zeta^2 q^i)}{h(x,y|\zeta,q)};\\\\

h(x,y|\zeta,q)&=& (1-\zeta^2 q^{2i})^2-(1-q)\zeta q^i (1+\zeta^2 q^{2i}) x y + (1-q)\zeta^2(x^2+y^2)q^{2i}.
\earr
\label{eq:biv-cqn}
\end{equation}
The $f_{CqN}$ and $f_{biv-qN}$ are normalized to 1 over the range $s(q)$, which can be inferred from the following property,
\begin{equation}
 \int_{s(q)} H_n(x|q) f_{CqN}(x|y;\zeta,q)\; dx = \zeta^n H_n(y|q).
 \label{eq:zfcqn}
\end{equation}
The first four moments of the $f_{CqN}$ can be given \cite{KM2020c} as,
\begin{equation}
\barr{rcl}
 \text{Centroid} &=& \zeta y,  \\\\ \text{Variance} &=& 1-\zeta^2 \;,\\\\

 \text{Skewness,}\;\gamma_1 &=& - \dis \frac{\zeta(1-q)y}{\sqrt{1-\zeta^2}}, \\\\ \text{Excess,} \; \gamma_2 &=& (q-1)+ \dis \frac{\zeta^2(1-q)^2 y^2+\zeta^2(1-q^2)}{(1-\zeta^2)}\;.
\earr
\label{eq:mom}
\end{equation}

Recently, it is shown that generating function for $q$-Hermite polynomials describes Gaussian to semi-circle transition in the eigenvalue density as $k$ changes from from $1$ to $m$ in spectral densities using $k$-body EGOE and their Unitary variants EGUE, both for fermion and boson systems \cite{Manan-Ko}. Very recently, in the strong coupling domain the lowest four moments of the strength function for $k$-body fermionic embedded ensemble are obtained and it is shown that they are essentially same as that of $f_{CqN}$ \cite{KM2020c}. Therefore, one can use $f_{CqN}$ distribution to represent the smooth forms of the strength functions and analyze the wavefunction structure in quantum many-body systems with $k$-body interactions. With this, the width of $f_{CqN}$ (and also of the strength fucntion) is related to  the correlation coefficient $\zeta$ by Eq.~\eqref{eq:mom}. In the next section, we will present our results for the variation of parameter $q$ and the correlation coefficient $\zeta$ as a function of $k$-body interaction strength $\lambda$ in  BEGOE(1+$k$). Also, a complete analytical description of $\zeta$, in terms of $N,m$,$k$ and $\lambda$, for BEGOE(1+$k$) is given.

\section{Parameter dependence of $q$ and $\zeta$ : results for BEGOE(1+$k$)}
\label{sec:3}

\subsection{Formula of $q$-parameter}
\label{sec:3a}

\begin{figure}[tbh!]
\centering
\includegraphics[width=0.6\textwidth,bb= 5 5 725 468]{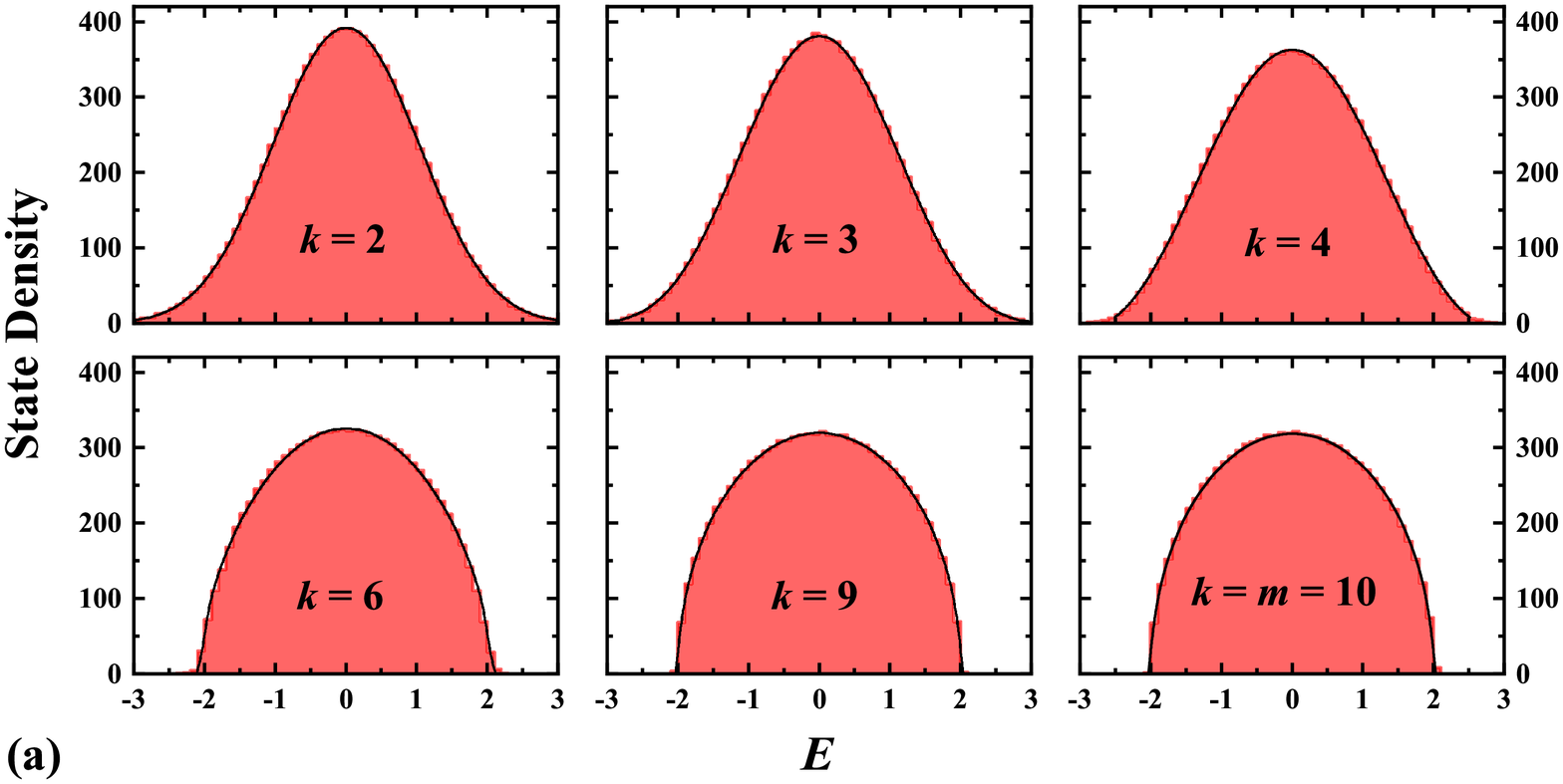}
\begin{tabular}{cc}
	\includegraphics[bb=10 0 370 305, width=0.44\textwidth]{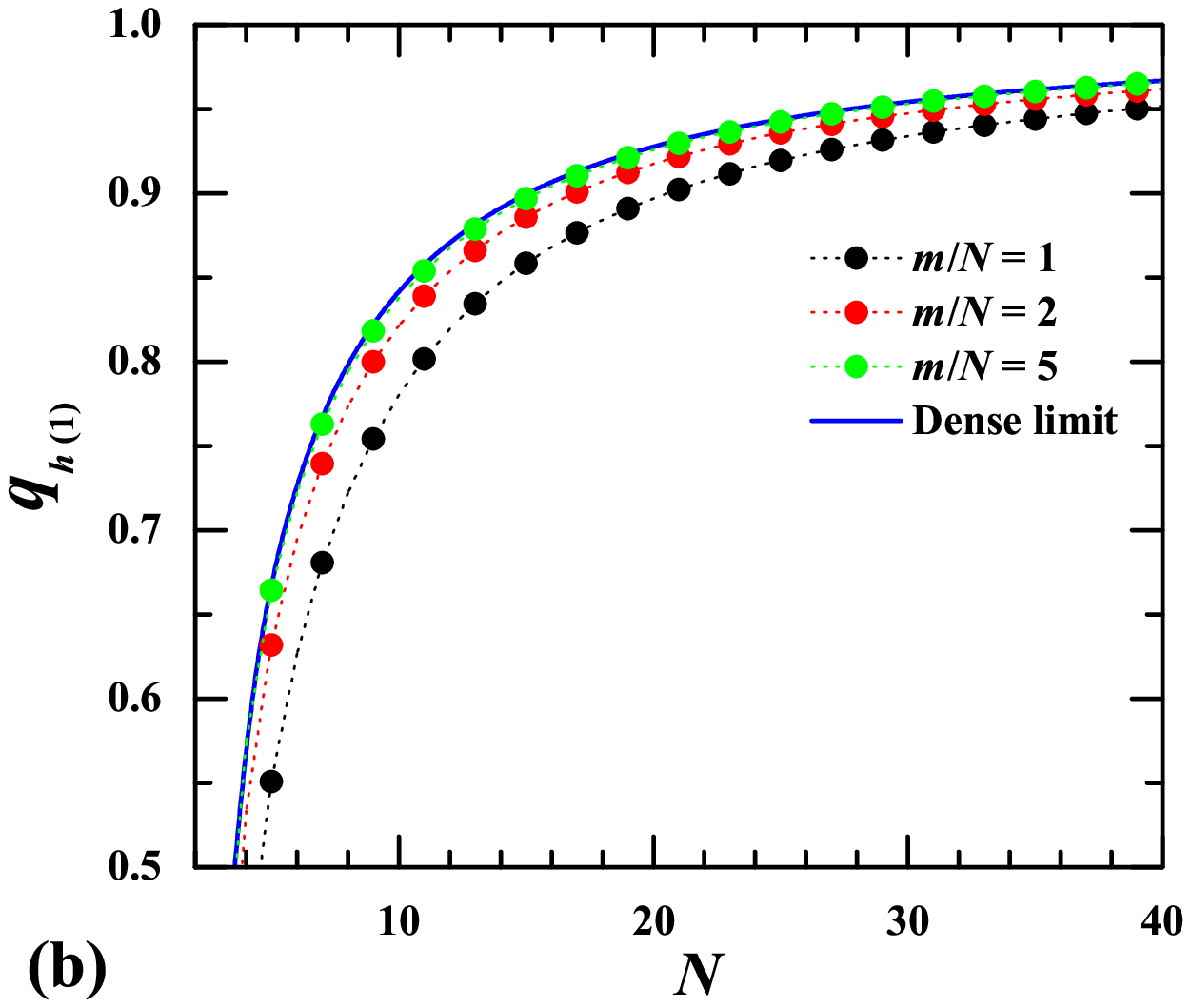}&\includegraphics[bb=10 5 360 310,width=0.43\textwidth]{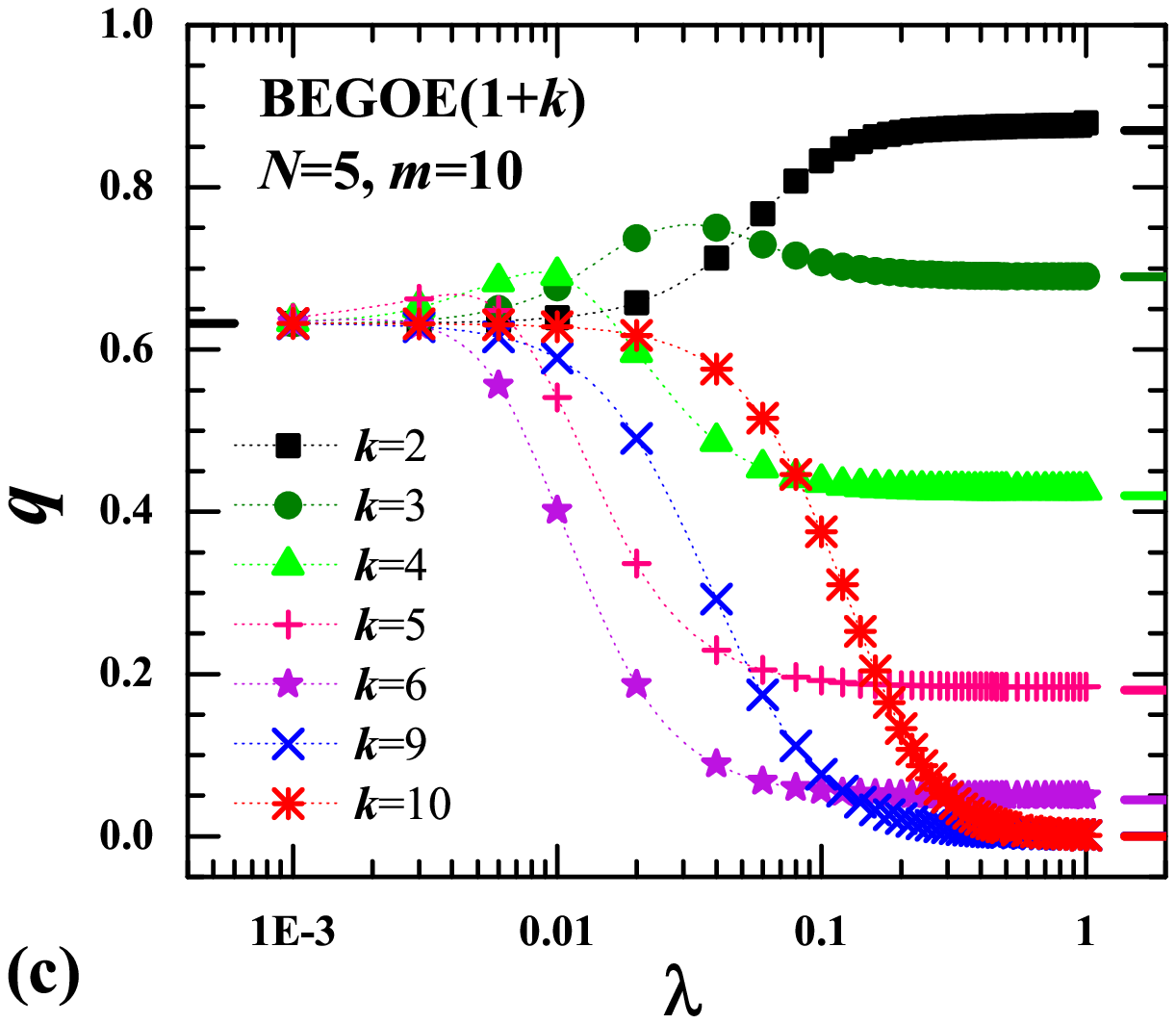}
\end{tabular}
\caption{(a) Histograms represent the state density vs. normalized energy $E$ results of the spectra of a 100 member BEGOE($1+k$) ensemble with $m = 10$ bosons in $N=5$ sp states for different $k$ values. The strength of interaction $\lambda = 0.5$ is chosen and in the plots $\int \rho(E) dE =d$. Ensemble averaged state density histogram is compared with $q$-normal distribution (continuous black curves) given by $f_{qN}(x|q)$ with the corresponding $q$ values given by Eq.~\eqref{eq.ent9}. (b) $q_{h(1)}$ vs. $N$ for various values of $m/N$. $q_{h(1)}$ is obtained using Eq.~\eqref{eq.q-h1} with sp energies $\epsilon_i=i+1/i$. Dense limit curve corresponds to the result with $m/N = 1000$. (c) Ensemble averaged $q$ vs. $\lambda$ for a 100 member BEGOE(1+$k$) ensemble with $m = 10$ bosons in $N=5$ sp states for different $k$ values. The horizontal black mark on left $q$-axis indicates $q$ estimate for $H=h(1)$ given by Eq.~\eqref{eq.q-h1}, while the colored marks on right $q$-axis represent the $q$ values, given by Eq.~\eqref{eq.ent9}, for corresponding $k$-body rank with $H=V(k)$. See text for more details.}\label{fig.rho-q}
\end{figure}

It has already been demonstrated that the state density for EE($k$)(and also EE(1+$k$)) in general exhibits Gaussian to semi-circle transition as $k$ increases from $1$ to $m$ \cite{MF}. This is now well verified in many numerical calculations and analytical proofs obtained via lower order moments \cite{kota-book,KC-18,kc-18b,Ag-02,BRW,SM}. Figure~\ref{fig.rho-q}(a) represents ensemble averaged state density obtained for a 100 member BEGOE(1+$k$) ensemble with $m = 10$ bosons distributed in $N = 5$ sp states and the body rank of interaction changing from $k$ = 2 to 10. In these  calculations, the eigenvalue spectrum for each member of the ensemble is first zero centered ($\epsilon_H$ is centroid) and scaled to unit width ($\sigma_H$ is width) and then the histograms are constructed. The results clearly display transition in the spectral density from Gaussian to semi-circle form as $k$ changes from 2 to $m=10$. With $E$ as zero centered and using $x = E/\sigma_H$, the numerical results are compared with the normalized state density $\rho(E) = d\;f_{qN}(x|q)$ with $\epsilon_H-\frac{2\sigma_H}{\sqrt{1-q}} \leq E \leq \epsilon_H+\frac{2\sigma_H}{\sqrt{1-q}}$. Here the parameter $q$ is computed using the formula, valid for BEGOE($k$)(i.e. $H=V(k)$), given in \cite{Manan-Ko},
\be
\barr{l}
q_{V(k)} \sim \dis\binom{N+m-1}{m}^{-1} \dis\sum_{\nu=0}^{\nu_{max}=\min[k,m-k]}\; \dis\frac{
X(N,m,k,\nu) \;d(g_\nu)}{\l[\Lambda^0(N,m,k)\r]^2} \;;\;\; \\\\
X(N,m,k,\nu) = \Lambda^\nu(N,m,m-k)\;\Lambda^\nu(N,m,k)\;;\\ \\
\Lambda^\nu(N,m,r) =  \dis\binom{m-\nu}{r}\;\dis\binom{N+m+\nu-1}{r}\;, \\\\
d(g_\nu)  = \dis\binom{N+\nu-1}{\nu}^2-\dis\binom{N+\nu-2}{\nu-1}^2\;.
\earr \label{eq.ent9}
\ee
In the strong coupling domain, one can also apply Eq.\eqref{eq.ent9} to BEGOE(1+$k$), as the $k$-body part of the interaction is expected to dominate over one-body part. One can see that the ensemble averaged results in Figure~\ref{fig.rho-q}(a) are in excellent agreement with the smooth forms obtained using $f_{qN}$. With $\lambda=0$ in Eq.\eqref{eq.ent2} i.e. one-body part $h(1)$ only, the analytical formula of $q$ for bosons, based on trace propagation method \cite{KP80}, can be given as,
\be
\barr{lcl}
q_{h(1)} &=& {\langle h(1)^4 \rangle}^m - 2 \\\\
 &=& \dis{ \{ \frac{3 (m-1) N (1+N) (1+m+N)}{m (2 + N) (3 + N) (m + N)}-2 \} }\\
 &&+ \dis{ \frac{m^2+(N+m)^2+(N+2m)^2}{m(N+m)} \frac{\sum_{i=1}^N \tilde{\epsilon_i}^4}{(\sum_{i=1}^N \tilde{\epsilon_i}^2)^2}}.\\
\earr
\label{eq.q-h1}
\ee
Here, ${\langle h(1)^4 \rangle}^m$ is the reduced fourth moment of one-body part and  $\tilde{\epsilon_i}$ are the traceless sp energies of $i$'th state. With $H=h(1)$ and uniform sp energies $\epsilon_i=i$, Eq.\eqref{eq.q-h1} gives $q=0.71$ for ($m=5,N=10$) and $q=0.68$ for ($m=10,N=5$). While with sp energies $\epsilon_i=i+1/i$, used in the present study, one obtains $q=0.68$ for ($m=5,N=10$) and $q=0.63$ for ($m=10,N=5$). Figure \ref{fig.rho-q}(b) shows variation of $q_{h(1)}$ as a function of $N$ for various values of $m/N$. Here, sp energies $\epsilon_i=i+1/i$ are used. It can be clearly seen that in the dense limit ($m\rightarrow\infty$, $N \rightarrow \infty$ and $m/N \rightarrow \infty$),  $q_{h(1)} \rightarrow 1$.  In the dilute limit ($m\rightarrow\infty$, $N \rightarrow \infty$ and $m/N \rightarrow 0$), similar variation in $q_{h(1)}$ can be observed due to $m \leftrightarrow N$ symmetry between the dense limit and the dilute limit as identified in \cite{Pa-00,KP80}. Furthermore, the variation of parameter $q$ is also studied as the interaction strength $\lambda$ varies in BEGOE(1+$k$) for a fixed body rank $k$. Here, the ensemble averaged value of $q$ is computed for a system of 100 member BEGOE(1+$k$) ensemble with $m=10$ bosons in $N=5$ sp states and results are shown in Figure \ref{fig.rho-q}(c). $q$ estimates are also shown in the figure by horizontal marks for $H=h(1)$ and $H=V(k)$ on left and right vertical axes respectively. One can see that for very small values of $\lambda$, ensemble averaged $q$ values are found very close to $q_{h(1)}$ for all body rank $k$. While for a sufficiently large $\lambda$, where $k$-body part dominates over one-body part and ensemble averaged $q$ values reach corresponding $q_{V(k)}$ given by Eq.\eqref{eq.ent9}. From the variation of ensemble averaged $q$ values in Figure \ref{fig.rho-q}(c), one can see that the shape of the state density takes intermediate form between Gaussian to semi-circle as $\lambda$ changes in BEGOE(1+$k$) for a fixed $k$. Therefore, the $q$-normal distribution $f_{qN}$ formula can be used to describe the transition in the state density with any value of $\lambda$ and $k$ in BEGOE(1+$k$).

\subsection{Formula of $\zeta$}
\label{sec:3b}

\begin{figure}[tbh!]
	\centering
		\includegraphics[width=\textwidth]{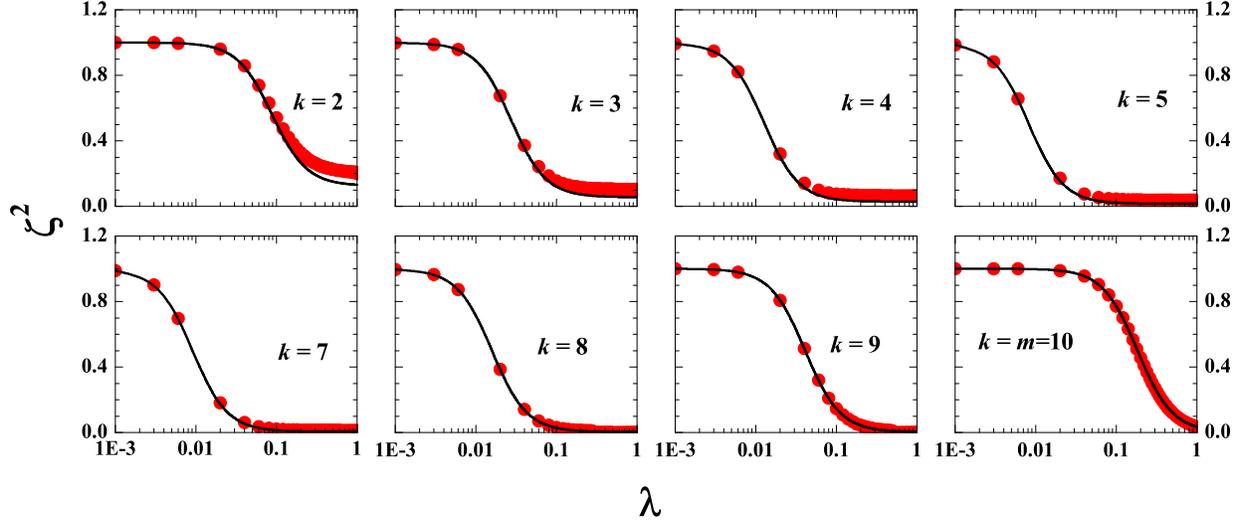}
	\caption{Ensemble averaged $\zeta^2$ (red solid circles) as a function of interaction strength $\lambda$, calculated for BEGOE(1+$k$) ensemble with $N = 5,m = 10$ example, are shown for different $k$ values. The smooth black curves are due to Eq.\eqref{eq.zeta2} using fixed sp energies $\epsilon_i=i+1/i$ employed in the present study.}
	\label{fig-zeta}	
\end{figure}

The parameter $\zeta$, which is the correlation coefficient between full Hamiltonian $H$ and the diagonal part $H_{\text{dia}}$ of the full Hamiltonian, is related to the width $\sigma_F$ of the strength functions, given by,
\be
\zeta=\sqrt{1-\dis\frac{\sigma_{H_{\text{off-dia}}}^2}{\sigma_H^2}} = \sqrt{1-\sigma_F^2},\; \; \;\; \sigma_F=\dis\frac{\sigma_{H_{\text{off-dia}}}}{\sigma_H}
\label{eq.zeta}
\ee
In the above equation, $\sigma_H^2$ and $\sigma_{H_{\text{off-dia}}}^2$ are variances of the eigenvalue distribution using full Hamiltonian and by taking all diagonal matrix elements as zero, respectively. Since $\zeta$ and $\sigma_F$ are simply related as $\sigma_F^2=1-\zeta^2$, here the discussion is in terms of $\zeta$. For BEGOE(1+$k$) ensemble, analytical expression for $\zeta$ based on the method of trace propagation can be derived as follows. For $H=V(k)$ i.e. with all sp energies as degenerate, it is known that \cite{Ag-02},
\be
\barr{rcl}
\sigma_{H=V(k)}^2 &=& \dis T(N,m,k) \binom{N+k-1}{k}^{-1} \;  \sum_{\alpha,\beta} \overline{ w^2_{\alpha \beta}}\;, \\\\
T(N,m,k)& =& \dis \Lambda^0(N,m,k)/ \binom{N+k-1}{k}\;.
\earr
\label{eq.sigvk}
\ee
Here, $\alpha$ and $\beta$ denote $k$-particle states. In $k$-particle space, the $H$ matrix is GOE. Therefore, the $k$-particle matrix elements $w_{\alpha \beta}$ are Gaussian random variates with zero mean and unit variance. The variance of diagonal matrix elements is $\overline{w^2_{\alpha \alpha}}=2$ while that of off-diagonal matrix elements is $\overline{w^2_{\alpha \beta}}=1$ for ($\alpha \ne \beta$). With this,
\be
\sigma_{H=V(k)}^2 = T(N,m,k) \; \binom{N+k-1}{k}^{-1} \left \{ 2 \times \text{no-dia} + 2 \times \text{no-offdia} \right \},
\ee
here the number of independent diagonal $k$-body matrix elements is 'no-dia'$= \binom{N + k - 1}{k}$ and that of off-diagonal is 'no-offdia'$= \frac{1}{2} \binom{N + k - 1}{k} \{\binom{N + k - 1}{k}-1 \}$. Similarly, $\sigma_{H_{\text{off-dia}}}$ is given by removing the contribution of diagonal $k$-body matrix elements from the above equation. Then using Eq.\eqref{eq.zeta} for $H=V(k)$,
\be
\zeta^2 = \frac{4}{{N+k-1 \choose k} + 1} \;.
\ee
Here, it can be immediately seen that $\zeta^2$ is independent of $m$ for BEGOE($k$). In the dense limit  with $N \rightarrow \infty$ and $m \rightarrow \infty$, $\sigma_F \rightarrow 1$ giving $\zeta \rightarrow 0$  as was suggested in \cite{Ch-03}. Also, with $k<<m$, $\zeta^2 \propto 1/N^k$. Using $m \leftrightarrow N$ symmetry between the dense limit and the dilute limit formula \cite{Pa-00,KP80}, we have $\zeta^2 \propto 1/m^k$ in the dilute limit and this result is in agreement with \cite{KM2020c}. Going further, with inclusion of one-body part defined by the external sp energies ($\epsilon_i$), and with $H = h(1) + \lambda V(k)$, we have
\be
\barr{rcl}
\sigma_{H}^2 &=& \sigma_{h(1)}^2 + \lambda^2 \; \sigma_{V(k)}^2,\\\\
&=& \frac{m(N+m)}{N(N+1)} \; \sum \tilde{\epsilon_i}^2 + \lambda^2\; \sigma_{V(k)}^2.
\earr
\ee
The analytical expression for $\zeta^2$ can be given by,
\be
\zeta^2=\frac{\frac{m(N+m)}{N(N+1)}\; \sum \tilde{\epsilon_i}^2 + 2\; \lambda^2\; T(N,m,k)} {\frac{m(N+m)}{N(N+1)}\; \sum \tilde{\epsilon_i}^2 + \lambda^2\; T(N,m,k)\; \{1 + \binom{N+k-1}{k}\} }\;.
\label{eq.zeta2}
\ee
In the above equation, the contribution from the diagonal part of $V(k)$ is also included into the numerator term. The analytical expression for $\zeta^2$ given by Eq.\eqref{eq.zeta2} is tested with the numerical ensemble averaged results obtained using a 100 member BEGOE(1+$k$) ensemble with $(m=10, N=5)$. The results of $\zeta^2$ as a function of $k$-body interaction strength $\lambda$ for different body rank $k$ are presented in Figure~\ref{fig-zeta}. The black smooth curve in each plot is obtained using Eq.\eqref{eq.zeta2} with fixed sp energies employed in the present study. It can be seen from the results that agreement between the ensemble averaged values (red solid circles) and the smooth forms obtained by Eq.\eqref{eq.zeta2} is very good for all $k$ values. Small difference with large $\lambda$, for $k<5$, is due to neglect of induced sp energies. The contribution of induced sp energies reduces as $\lambda$ and $k$ increases. One can see from the results shown in Figure~\ref{fig-zeta} that the width of the strength function is strongly dependent on $\lambda$. For $\lambda \rightarrow 0$, $\zeta^2 \rightarrow 1$ for all $k$ and the strength functions are known to be represented by $\delta$ functions. With increase in $\lambda$ i.e.$\lambda \ge \lambda_C$, the strength functions are known to be described by the Briet-Wigner (Lorentz) form. With further increase in $\lambda >> \lambda_F$,  $\zeta^2$ goes on decreasing smoothly leading to a fully chaotic domain giving the Gaussian or semi-circle or intermediate to Gaussian and semi-circle  character of the strength functions depending upon the values of $\lambda$ and $k$.  One can also observe the BW to Gaussian to semi-circle transition in strength functions by changing both $\lambda$ and $k$. Therefore, it is possible to have a shape intermediate to BW and semi-circle for some values of $\lambda$ and $k$ \cite{LBBZ}. 

\begin{figure}[tbh!]
	\centering
	\includegraphics[width=0.5\textwidth]{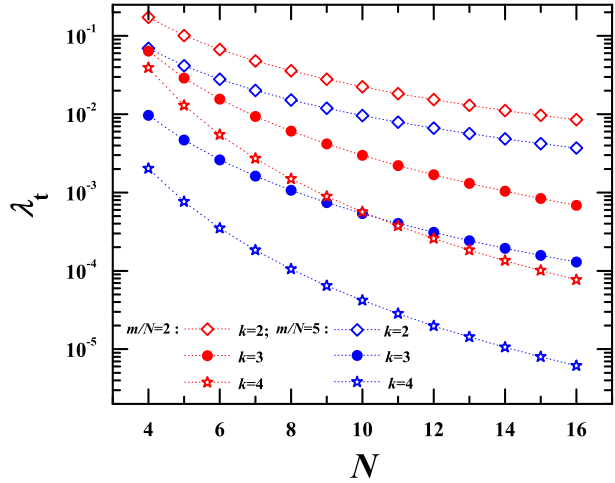}
	\caption{Variation of marker $\lambda_t$  as a function of $N$ for dense boson systems with BEGOE(1+$k$). Results are shown for various values of ($k$,$m/N$) using Eq.\eqref{eq.lt}.}
	\label{fig-lt}	
\end{figure}

For two-body interaction, the thermodynamic region $\lambda=\lambda_t$ can be determined using the condition $\zeta^2 = 0.5$ \cite{Ch-PLA,angam}; i.e. the spreading produced by one-body part and two-body part are equal. Similarly, one can obtain marker $\lambda_t$ for $k$-body interactions in presence of mean field by considering the spreading produced by one-body part and $k$-body part equal in Eq.\eqref{eq.zeta2}. Solving it for $\lambda$, ($m$, $N$, $k$) dependence of marker $\lambda_t$ is given by,
\be
\lambda_t=\sqrt{\frac{m (N+m)\;\sum \tilde{\epsilon_i}^2 }{ N(N+1)\Lambda^0(N,m,k)(1-3\;\binom{N+k-1}{k}^{-1})}}\;\;.
\label{eq.lt}
\ee
Figure \ref{fig-lt} shows the variation of marker $\lambda_t$ in dense boson systems with BEGOE(1+$k$) as a function of $N$ for the fixed sp energies used in the present study. The results are shown for body rank values $k=2,3$ and $4$, and with $m/N=2$ and $5$. From the results one can clearly see that $\lambda_t$ decreases as the rank of the interaction $k$ increases. Hence, the thermalization sets in faster as the rank of interaction $k$ increases.

Recently, using $k$-body embedded ensembles both for fermions and bosons, it is demonstrated that in the thermalization region ($\lambda \geq \lambda_t$), shape of the strength functions changes from Gaussian to semi-circle for the states close to the center of the spectrum as the rank of the interaction $k$ increases and they can be well represented by $f_{qN}$ form for all $k$ values in $V(k)$ \cite{Manan-Ko}. The strength functions are symmetrical in $E$ near the center of the spectrum as is the result with $f_{qN}$. However, it is seen in some calculations with $k=2$ that the strength functions become asymmetrical in $E$ as one moves away from the center \cite{CK2017}. This feature can be incorporated by representing strength function using $f_{CqN}$ which can not be generated by $f_{qN}$. This will be verified with a numerical example in the next section and more importantly, a single interpolating function $f_{CqN}$, in terms of parameters $q$ and $\zeta$, is considered for describing Gaussian to semi-circle transition in the strong coupling domain as the body rank $k$ in BEGOE(1+$k$) is changed.

\section{Strength function}
 \label{sec:4}
\begin{figure}[tbh!]
	\centering
	\begin{tabular}{c}
		\includegraphics[width=0.35\textwidth,bb= 5 0 450 660]{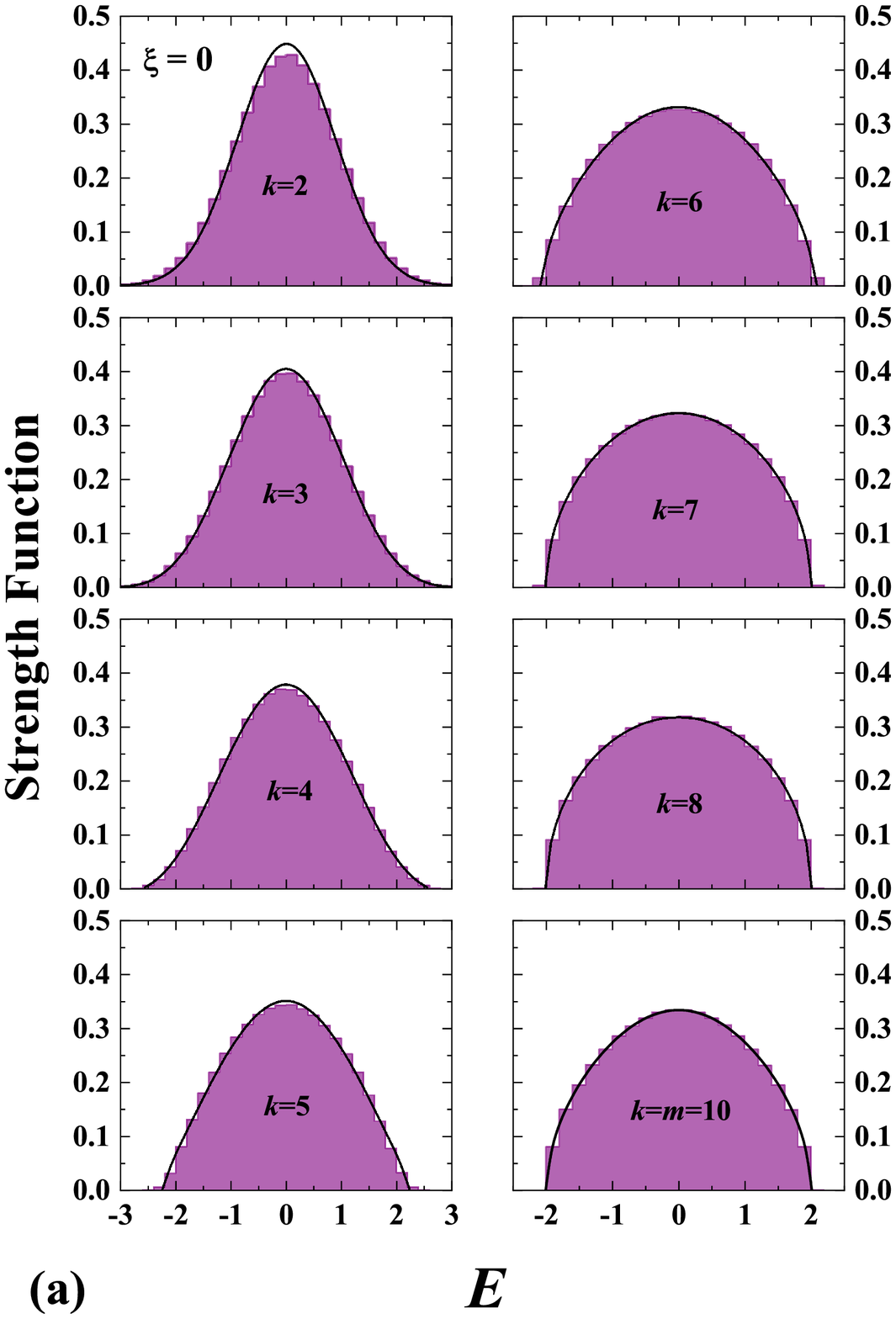}
	\end{tabular}
	\begin{tabular}{cc}
		\includegraphics[width=.35\textwidth,bb= 5 0 450 660]{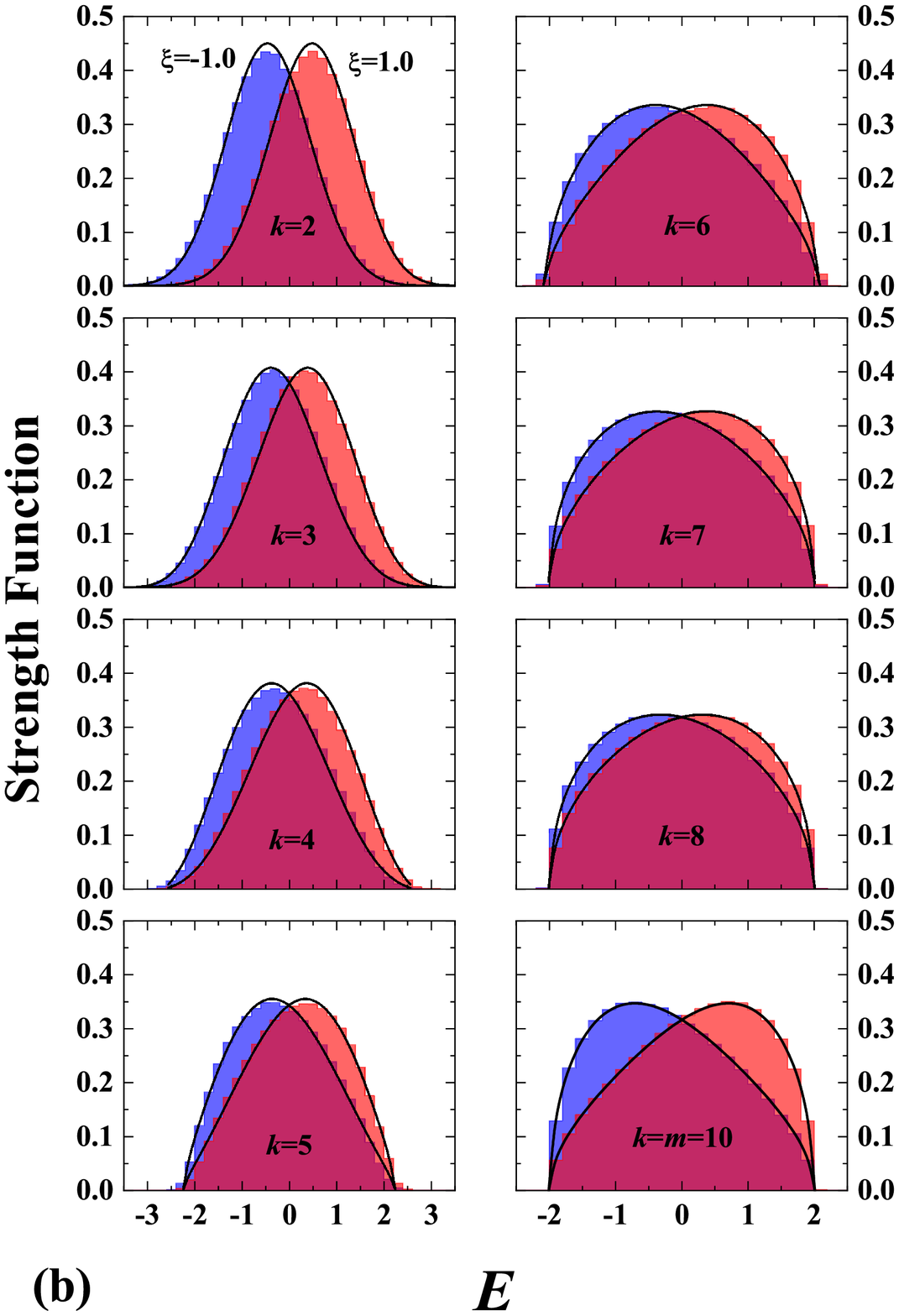}&\includegraphics[width=.35\textwidth,bb= 5 0 450 660]{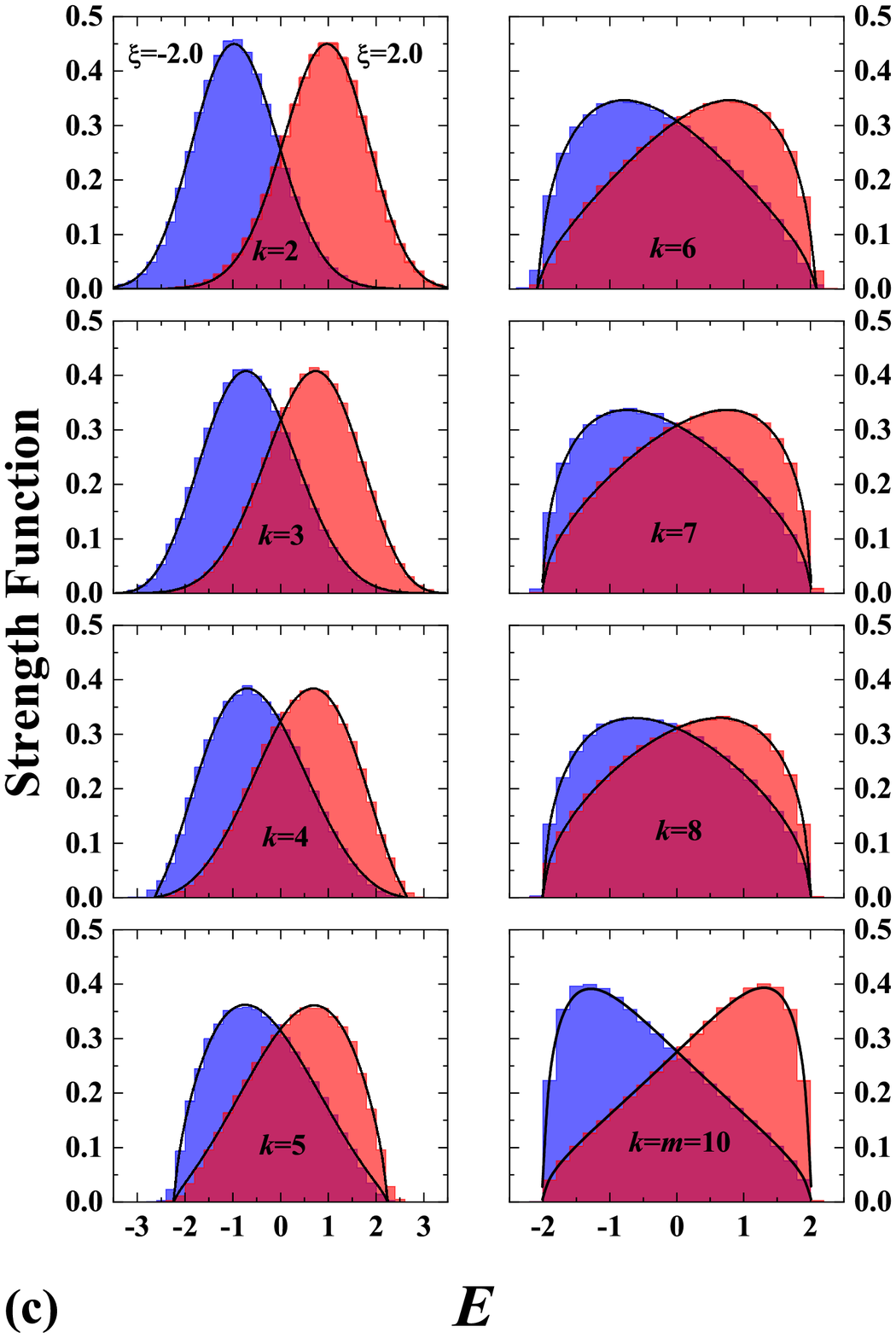}
	\end{tabular}
\caption{Strength function vs. normalized energy $E$ for a system of $m = 10$ bosons in $N = 5$ sp states with $\lambda=0.5$ for different $k$ values in BEGOE(1+$k$) ensemble. An ensemble of 250 members is used for each $k$. Strength function plots are obtained for (a) $\xi=0$ (purple histogram) , (b) $\xi=-1.0$ (blue histogram) and $1.0$ (red histogram) and (c) $\xi=-2.0$ (blue histogram) and $2.0$ (red histogram). In the plots $\int F_{\xi}(E) dE=1$. The continuous black curves are due to fitting with $f_{CqN}$ given by Eq.~\eqref{eq.fcqn} using $q$ and $\zeta$ values obtained by Eq.~\eqref{eq.ent9} and Eq.~\eqref{eq.zeta}, respectively. See text for more details.}
\label{fig.fke}
	\end{figure}

\begin{figure}[tbh!]
	\centering
	\begin{tabular}{ccc}
		\includegraphics[width=0.3\textwidth,bb= 5 5 396 320]{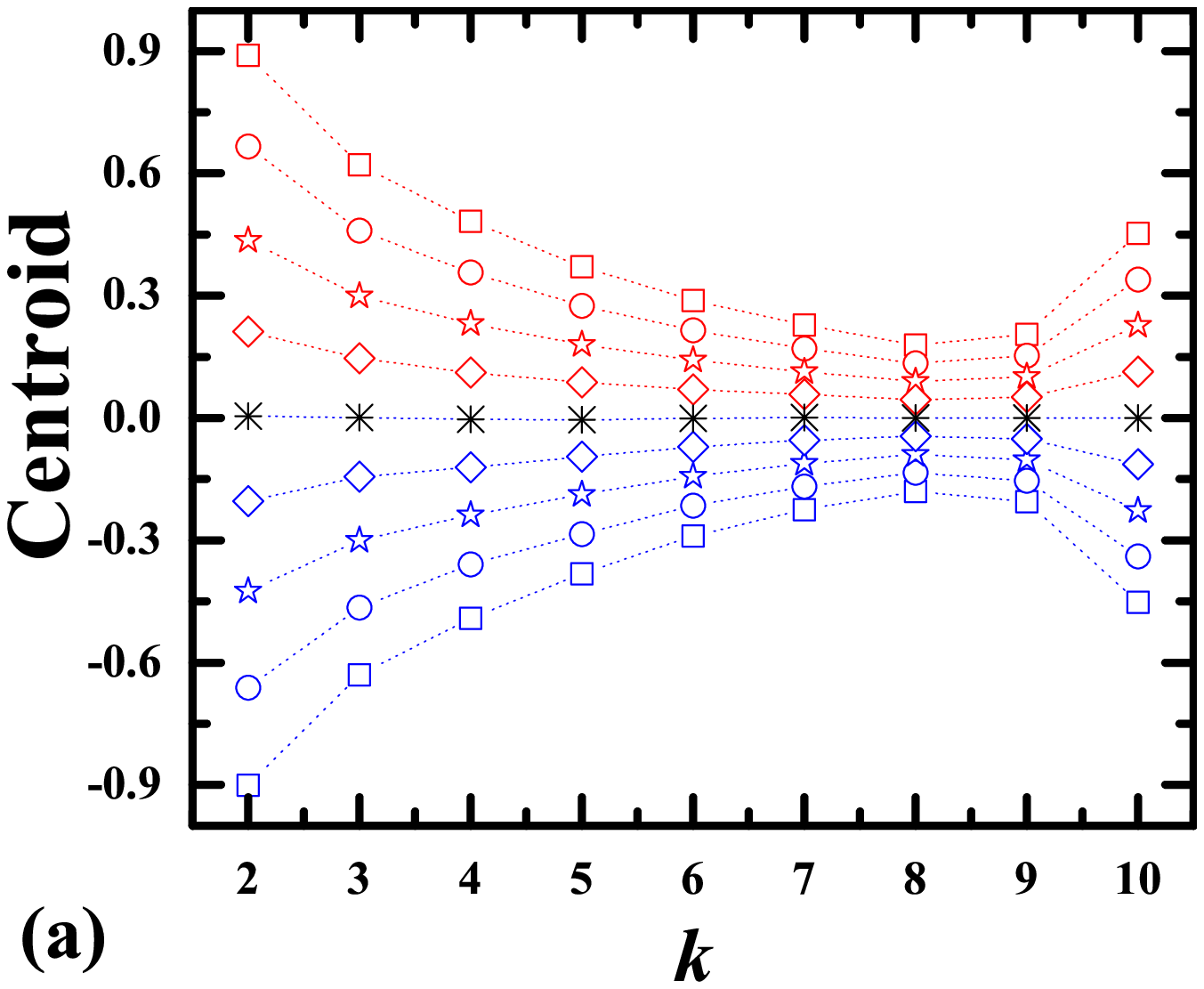}&
		\includegraphics[width=0.3\textwidth,bb= 5 5 396 320] {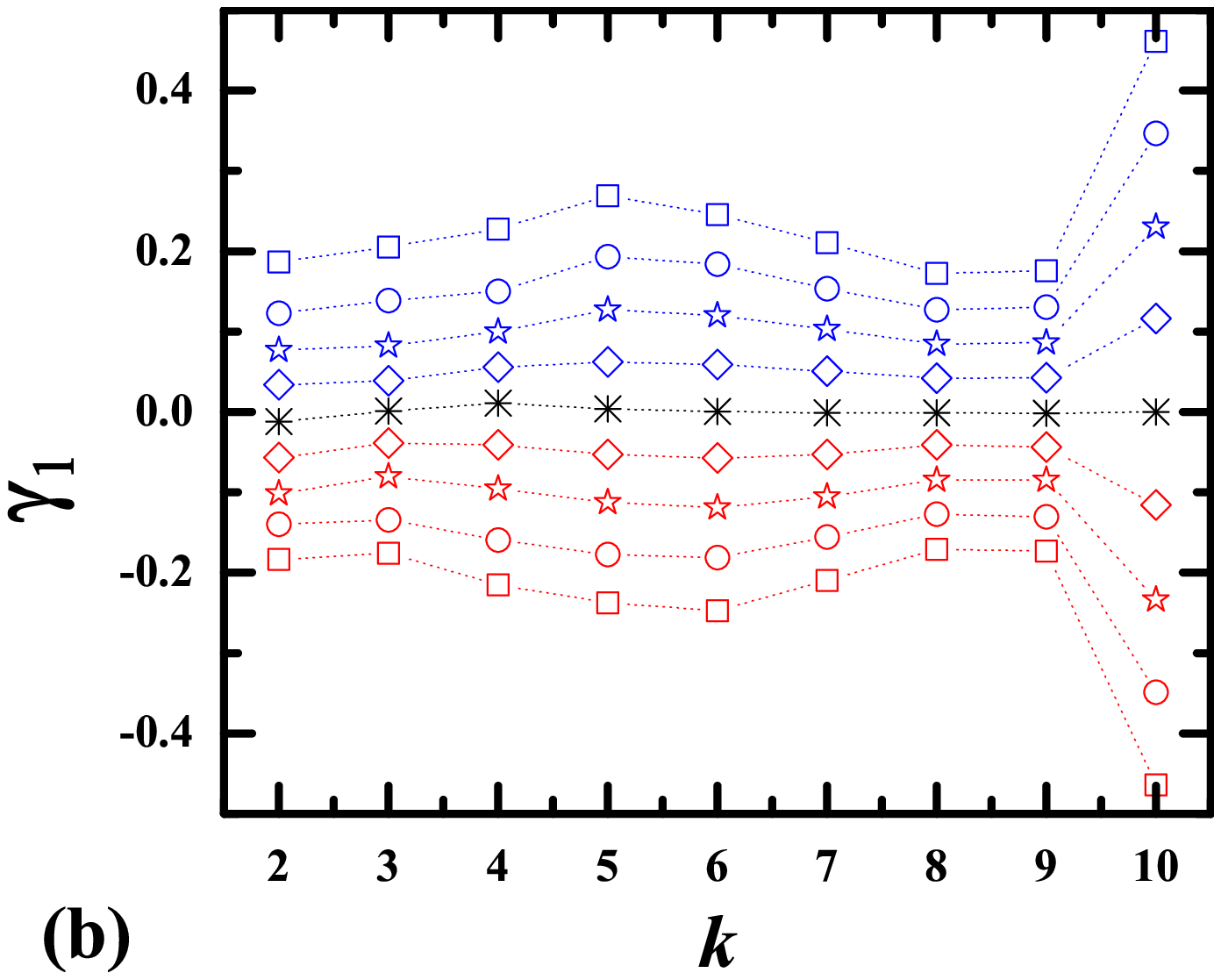}& \includegraphics[width=0.3\textwidth,bb= 5 5 396 320] {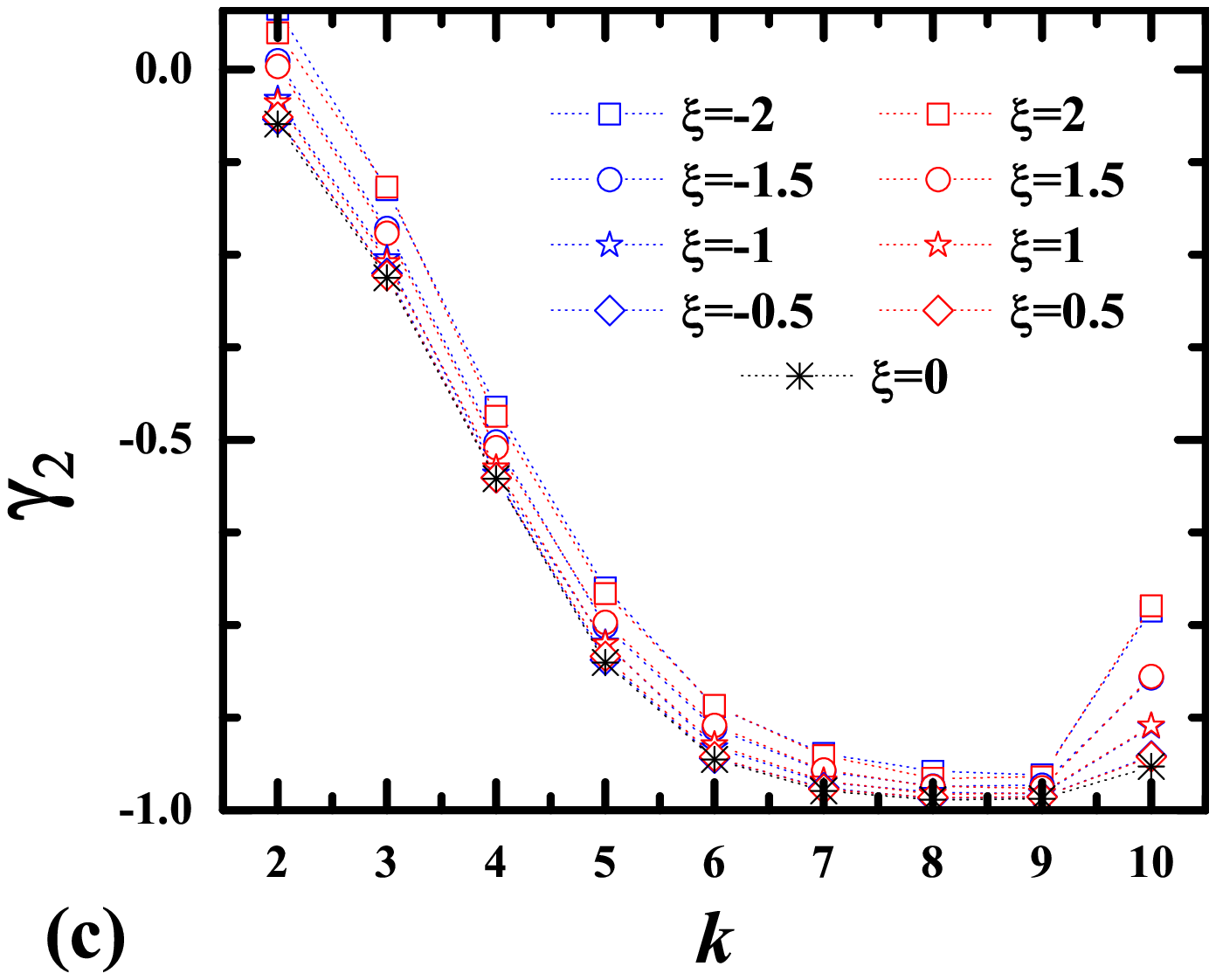}
	\end{tabular}
 	\caption{Ensemble averaged (a) Centroid, (b) $\gamma_1$ and (c) $\gamma_2$ as a function of body rank $k$ for the strength function results presented in Figure \ref{fig.fke}. Results are shown for various values of $\xi$.}
  	\label{fig-mom-fcqn}	
 \end{figure}

Given $m$-particle basis state  $\l| \kp \ran$, the diagonal matrix elements of $m$-particle Hamiltonian $H$ are denoted as energy $\xi_{\kp}$, so that $\xi_{\kp}=\langle \kp | H | \kp \rangle$. The diagonalization of the full matrix $H$ gives the eigenstates $\l| E_i \ran$ with eigenvalues $E_i$, where $\l| \kp \ran = \sum_i C_{\kp}^{i} \l| E_i \ran $. The strength function that corresponds to the state $\l| \kp \ran$ is  defined as $ F_{\xi_\kp}(E) =\sum_i {|C_{\kp}^{i}|}^2 \;\delta(E - E_i)$. In the present study,  we take the $\l| \kp \ran$ states to be the eigenstates of $h(1)$. In order to get an ensemble averaged form of the strength functions, the eigenvalues $E_i$ are scaled to have zero centroid and unit variance for the eigenvalue distribution. The $\kp$-energies, $\xi_{\kp}$, are also scaled similarly. Now, for
each member, all ${|C_{\kp}^{i}|}^2$ are summed over the basis states
$\kp$ with energy $\xi$ in the energy window
$\xi \pm \Delta$. Then, the ensemble averaged $F_{\xi}(E)$ vs.
$E$ are constructed as histograms by applying the normalization condition
$\int_{s(q)} F_{\xi}(E)\;dE=1$.  In Figure~\ref{fig.fke}, histograms represent ensemble averaged $F_\xi(E)$ results for all body rank $k$ values with $\lambda=0.5$ using a 250 member BEGOE(1+$k$) ensemble with $m=10$ and $N=5$ system. The strength function plots are obtained for $\xi = 0.0, \pm 1.0$ and $\pm 2.0$.  The value of $k$-body interaction strength is chosen such that $\lambda >> \lambda_t$, i.e. the system exists in the region of thermalization \cite{kc-18b,Ch-PLA}. The histograms, representing BEGOE(1+$k$) results of strength functions, are compared with the conditional $q$-normal density function as given by,
\be
F_\xi(E)= f_{CqN}(x=E|y=\xi;\zeta,q).
\label{eq.fcqn}
\ee
The smooth black curves in Figure~\ref{fig.fke} for each $k$ are obtained via Eq.\eqref{eq.fcqn} using corresponding ensemble averaged $\zeta$ and $q$ values. With $\lambda >> \lambda_t$, $\zeta^2 << 1/2$, the $q$ value in Eq.\eqref{eq.fcqn} can fairly be given by Eq.\eqref{eq.ent9} \cite{KM2020c}. The results in Figure~\ref{fig.fke} clearly show very good agreement between the numerical histograms and continuous black curves for all body rank $k$. The $F_\xi(E)$ results for $\xi=0$ are given in Figure~\ref{fig.fke}(a) which clearly demonstrate that the strength functions are symmetric and also exhibit a transition from Gaussian form to semi-circle as $k$ changes from $2$ to $m=10$. The smooth form given by Eq.\eqref{eq.fcqn} using the conditional $q$-normal density function interpolates this transition very well. Going further, $F_\xi(E)$ results for $\xi \neq 0$ are shown in Figures~\ref{fig.fke}(b) and \ref{fig.fke}(c). One can see that $F_\xi(E)$ results are asymmetrical about $E$ as demonstrated earlier \cite{CK2017}. Also, $F_\xi(E)$ are skewed more in the positive direction for $\xi >0$ and skewed more in the negative direction for $\xi < 0$ and their centroids vary linearly with $\xi$. We have also computed the first four moments (centroid, variance, skewness ($\gamma_1$) and excess ($\gamma_2$)) of the strength function results shown in Figure~\ref{fig.fke} for the body rank $k$ going from $2$ to $m=10$. Figure \ref{fig-mom-fcqn} represents results for centroid, $\gamma_1$ and $\gamma_2$ for various values of $\xi$. As discussed earlier in Section \ref{sec:3}, the variance of the strength functions is independent of $\xi$ and simply related to correlation coefficient; for more details, see results of $\zeta^2$ (Figure~\ref{fig-zeta}). From the numerical results obtained for strength functions (Figure \ref{fig.fke}) along with results of lower order moments (Figure~\ref{fig-mom-fcqn}), one can clearly see that in the thermodynamic domain, the strength functions of dense interacting many-boson systems, with $k$-body interaction, follow the conditional $q$-normal distribution $f_{CqN}$. The results are also consistent with the analytical forms derived in \cite{KM2020c}.

In the study of thermalization and relaxation dynamics of an isolated finite quantum system after a random interaction quench, strength functions play an important role. Having tested that in the thermodynamic region with $\lambda >> \lambda_t$, ensemble averaged strength functions of dense boson systems with $k$-body interaction can be represented by smooth forms given by $f_{CqN}$, we will now utilize these interpolating forms, in the coming sections, to study fidelity decay and NPC in dense boson systems with $k$-body interaction.


\section{Fidelity decay after an interaction quench}
\label{sec:5}

\begin{figure}[tbh!]
\centering
\includegraphics[width=\textwidth]{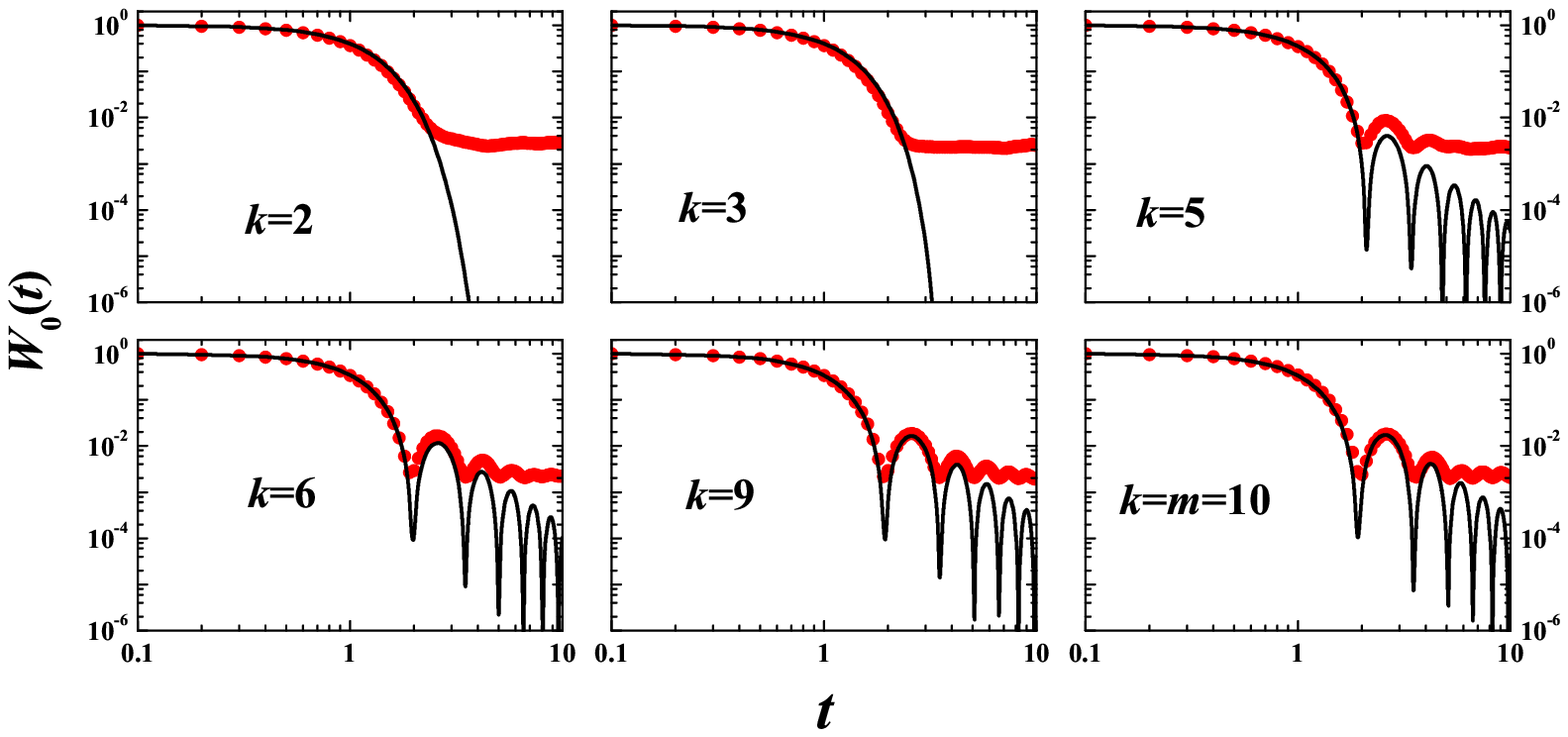}
\caption{Fidelity decay $W_0(t)$ as a function of time for a 100 member BEGOE(1+$k$) ensemble with $N = 5$ and $m = 10$ represented by the red solid circles; the $\psi(0)$ here corresponds to middle states of $h(1)$ spectrum. \textcolor{red}{ Here $t$ is measured in the units of $\sigma_H^{-1}$.} The black smooth curves are obtained by taking numerical Fourier transform of the strength functions represented by Eq.\eqref{eq.fcqn}.}
\label{fig:w0}
\end{figure}

Fidelity decay or return probability of a quantum system after a sudden quench is an important quantity in the study of relaxation of a complex (chaotic) system to an equilibrium state. Let's say the system is prepared in one of the eigenstates ($\psi(0)=\l| \kp \ran$) of the mean-field Hamiltonian $H=h(1)$. With the quench at $t=0$ by $\lambda V(k)$, the system evolves unitarily with respect to $H \rightarrow h(1)+\lambda V(k)$ and the state changes after time $t$ to $\psi(t)=\l| \kp(t) \ran=\exp(-iHt) \l| \kp \ran$. Then, the probability to find the system in it's initial unperturbed state after time $t$, called fidelity decay, is given by,
 \be
 \barr{lll}
W_{0}(t) &=& |\lan\psi(t)|\psi(0)\ran|^2 = \l| \sum_E \l[C_k^E\r]^2 \exp -iEt \r|^2 \\\\
&=& \int F_\xi(E) \exp-iEt\;dE \\\\ &=& \int_{s(q)} f_{CqN}(E|\xi;\zeta,q) \exp-iEt\;dE \;.
 \earr
\label{eq.w0}
\ee
Thus, fidelity is the Fourier transform in energy of the strength function; this is valid for times not very short or very long. In the thermalization region, the form of $F_\xi(E)$ is Gaussian for $k=2$ while it is semi-circle for $k=m$. These two extreme situations are recently studied, both analytically \cite{Ha-16} as well as numerically \cite{Lea,Lea-a,Lea-b}. The formula for $W_0(t)$ can be given in terms of width of $\lambda V(k)$ scaled by $\sigma_H$.  Clearly, following the results of the previous section, $f_{CqN}$ can be used to obtain $W_0(t)$ generated by BEGOE(1+$k$). As analytical formula for the Fourier transform of $f_{CqN}$ is not available, therefore we evaluated  Eq.\eqref{eq.w0} numerically. Figure~\ref{fig:w0} shows results for $W_0(t)$ (red solid circles) for a 100 member BEGOE(1+$k$) ensemble with $m=10$, $N=5$ and $\lambda=0.5$ for various $k$ values and they are compared with numerical Fourier transform (black smooth curves) of Eq.\eqref{eq.fcqn}. Here, we have used normalized eigenenergies  in the computation of $W_0$ and therefore the time $t$ is measured in the units of $1/\sigma_H$. It is clear from the results that the Fourier transform of $f_{CqN}$ describes the short-time behavior nicely and also captures the positions of the oscillations. The results generated here are consistent with the reported results  in \cite{Manan-Ko}, obtained using $f_{qN}$ form for $F_\xi(E)$.
	
It is known that in the strong interaction domain, the decrease in $W_0$ (for $k=2$) follows quadratic in time and this Gaussian decrease can last for a quite large time and after that an exponential one emerges \cite{lea-2019}. The transition time depends on the ratio of the spectral width and the square of the second moment of strength fucntion ($\sigma_F^2$). As here $\lambda >> \lambda_t$, $\zeta^2 \rightarrow 0$ giving $\sigma_F^2 \approx 1$, $t$ is in $1/\sigma_H$ units and the spectral width will be in $\sigma_H$ units. Therefore, the results in Figure~\ref{fig:w0} describe $W_0$ nicely for short time and the standard exponential decrease for long time for $k=2$ seems absent. The long time behavior of fidelity decay is of great interest as it is expected that $W_0$ surely demonstrates a power-law behavior i.e. $W_0(t) \propto t^{-\gamma}$ with $\gamma \geq 2$ implying thermalization \cite{Lea-power}, no matter how fast
the decay may initially be. As shown in \cite{Lea-power}, the power-law behavior appears due to the fact that the energy spectrum is bounded from both the ends. This condition is essentially satisfied by $f_{CqN}$. Therefore, it is important to analyze the long-time behavior of fidelity decay for embedded ensembles first to establish it's universality and second to test whether it can be explained with the use of $f_{CqN}$. These are open questions.

In the study of fidelity decay, strength function with $\xi =0$ is involved. However, the statistical properties, related to  wavefunction structure, namely NPC and $S^{\text{info}}$ can be written as integrals involving strength functions over all $\xi$ energies. Very recently, an integral formula for NPC in the transition strengths from a state as a function of energy for fermionic EGOE($k$) using the bivariate $q$-normal form is presented in \cite{KM2020}. In the past, the smooth forms, for NPC and $S^{\text{info}}$, were derived in terms of energy and correlation coefficient $\zeta$ for two-body interaction \cite{KS2001}. In the next section, we present our results for NPC and $S^{\text{info}}$ using  $f_{CqN}$ forms for the strength functions and compare with those for dense interacting boson systems with $k$-body interaction. 

\section{NPC and Information entropy}
\label{sec:6}

The NPC in wavefunction characterizes various layers of chaos in interacting particle systems \cite{FI2,PGS1998,CMejia1998} and for a system like atomic nuclei, NPC for transition strengths is a measure of fluctuations in transition strength sums \cite{KM2020}. For an eigenstate $|E_i \rangle$ spread over the basis states $|\kp \rangle$, with energies $\xi_\kp = \langle \kp|H|\kp \rangle$,
NPC (also known as inverse participation ratio) is defined as,
\be
\mbox{NPC}(E) = \left\{ {\dis\sum\limits_\kp {\left| {C_{\kp}^{i} } \right|^4 } } \right\}^{ - 1}
\label{eq.ijm-chv7}
\ee
NPC essentially gives the number of basis states $\l.\l|\kp\r.\ran$ that constitute an eigenstate with energy $E$. The GOE value for NPC is $d/3$. NPC can be studied by examining the general features of the strength functions $F_{\xi}(E)$. The smooth forms for NPC$(E)$ can be written as \cite{KS2001},
\be
\mbox{NPC}(E)= \dis\frac{d}{3} \left\{ \dis\int d\xi \; \dis\frac{\rho^{H_\kp}(\xi) [F_{\xi}(E)]^2}{[\rho^H(E)]^2} \right\}^{-1}\;,
\label{eq.ijm1}
\ee
where $\rho^{H_\kp}(\xi)$ and $\rho^{H}(E)$ are normalized eigenvalue densities generated by diagonal  Hamiltonian $H_\kp$ matrix and full Hamiltonian $H$ matrix, respectively. Taking $E$ and $\xi$ as zero centered and scaled by corresponding widths, the above equation can be written in terms of $f_{qN}$ and $f_{CqN}$ \cite{KM2020,KM2020c},

\be
\mbox{NPC}(E) = \dis\frac{d}{3} \left\{ \dis\int_{S(q)} d\xi \; \dis\frac{f_{qN}(\xi|q) [f_{CqN}(E|\xi;\zeta,q)]^2}{f_{qN}(E|q)} \right\}^{-1}\;,
\label{eq.ijm2}
\ee
In general, $q$'s in the above equation need not be same \cite{KM2020,KM2020c}. However, in the thermalization region, with $\zeta^2 \le 1/2$, one can approximate $\gamma_2 \approx (q-1)$ in Eq.\eqref{eq:mom}. Then, the formula for $q$ given by Eq.\eqref{eq.ent9} is valid for $f_{qN}$ as well as for $f_{CqN}$. This is well verified numerically in Section~\ref{sec:2}. Also, the results of $\gamma_2$ in Figure \ref{fig-mom-fcqn}(c) corroborate this claim. With this, it is possible to simplify Eq.\eqref{eq.ijm2} using Eqs.\eqref{eq:biv-cqn} and \eqref{eq:zfcqn} and a simple two parameter formula, valid in chaotic domain, for NPC can be written as,
\be
\mbox{NPC}(E) = \dis\frac{d}{3} \dis \left\{ \sum_{n=0}^{\infty} \frac{\zeta^{2n}}{[n]_q!}\, H_n^2(E|q) \right\}^{-1} ,
\label{eq.npc}
\ee
It is easy to see from above formula that NPC($E$) approaches GOE value $d/3$ as $\zeta \rightarrow 0$. Also for $q \rightarrow 1$, $f_{qN}$ and $f_{CqN}$ in  Eq.\eqref{eq.ijm2} reduce to Gaussian and then Eq.\eqref{eq.npc} gives similar results obtained for $k=2$ in \cite{KS2001}. We have tested this formula with numerical ensemble averaged BEGOE(1+$k$) results. Figure~\ref{fig-npc}, shows results for ensemble averaged NPC vs. normalized energy, for a 100 member BEGOE(1+$k$) with $m=10$ and $N = 5$ example for different values of $\lambda$ and $k$. The ensemble averaged NPC values are shown with red solid circles and continuous lines are obtained using the theoretical expression given by Eq.~\eqref{eq.npc}. One can see from the results that with fixed $k$ (i) for small value of $\lambda$, where the one-body part of the interaction is dominating, the numerical NPC values are zero and the theoretical curve is far away
from the numerical results indicating that the wavefunctions are completely localized (the bottom panels in Figure~\ref{fig-npc}); (ii) with further increase in $\lambda$,  the theoretical estimate for NPC in the chaotic domain is
much above the ensemble averaged curve indicating that the chaos has
not yet set in; (iii) However, with sufficiently large $\lambda$, we see that the ensemble averaged
curve is matching with the theoretical estimate given by Eq.~\eqref{eq.npc}, indicating that system is in chaotic domain corresponding to the thermalization region given by $\zeta^2 \sim 1/2$ \cite{Ch-PLA} and the strength functions $F_\xi(E)$ are well represented by conditional $q$ normal distribution. Again with further increase in $\lambda$ (the top panels in Figure \ref{fig-npc}), the match between the theoretical chaotic domain estimate and the ensemble averaged values is very well in the bulk part of the spectrum ($|E|<2$) for all values of $k$ with deviations near the spectrum tails. Hence, in the chaotic domain, the energy variation of NPC($E$) using Eq.~\eqref{eq.npc} is essentially given by two parameters, $\zeta$ and $q$. The results clearly show that the thermalization sets in faster with increase in the body rank $k$.

Another statistical quantity normally considered is the information entropy defined by
$S^{\text{info}} (E) =  -\sum_{\kp} p_\kp^i \ln p_\kp^i = - \sum_\kp |C_{\kp}^{i}|^2 \ln |C_{\kp}^{i}|^2$, here $p_\kp^i$ is the probability of basis state $\kp$ in the eigenstate at energy $E_i$. The localization length, $l_H$ is related to $S^{\text{info}}(E)$ by $l_H(E)=\exp{S^{info} (E)}/(0.48 d)$. Then the corresponding embedded ensemble expression for $l_H$ involving $F_\xi(E)$, can be written as\cite{KS2001},

\be
l_H(E)= - \dis\int d\xi \; \dis\frac{F_{\xi}(E)\; \rho^{H_\kp}(\xi)}{\rho^H(E)} \ln \left\{\dis\frac{F_{\xi}(E)}{\rho^H(E)}\right\}\;.
\label{eq.lh1}
\ee
Replacing $\rho^{H_\kp}(\xi)$ and $\rho^H(E)$ by $f_{qN}$ and $F_{\xi}(E)$ by $f_{CqN}$, formula for $l_H$ valid in chaotic domain is given by,
\be
l_H(E)= - \dis\int d\xi \; \dis\frac{f_{CqN}(E|\xi;\zeta,q) f_{qN}(\xi|q) }{f_{qN}(E|q)} \; \ln \left\{\dis\frac{f_{CqN}(E|\xi;\zeta,q)}{f_{qN}(E|q)} \right\}\;.
\label{eq.lh}
\ee
Simplifying Eq.\eqref{eq.lh} for $l_H$ is an open problem and therefore, it is evaluated numerically and results are compared with ensemble averaged numerical results of BEGOE(1+$k$). Figure~\ref{fig-lh}, shows results for ensemble averaged $l_H$ vs. normalized energy $E$ for a 100 member BEGOE($1+k)$ with $m=10$ bosons in $N = 5$ sp states for different values of $k$. Here, we choose $k$-body interaction strength $\lambda=1$ so that the system will be in thermalization region. Numerical embedded ensemble results (red solid circles) are compared with theoretical estimates (black curves) obtained using Eq.~{\eqref{eq.lh}. The $\zeta$ values are shown in the figure. A very good agreement between numerical results and smooth form is obtained for all values of $k$ in the bulk of the spectrum with small deviations near the spectrum tails. Hence, in the chaotic domain, the energy variation of $l_H(E)$, with Eq.~\eqref{eq.lh}, is essentially given by conditional $q$ forms for the strength functions.

 \begin{figure}[!tbh]
\centering
\includegraphics[width=\textwidth]{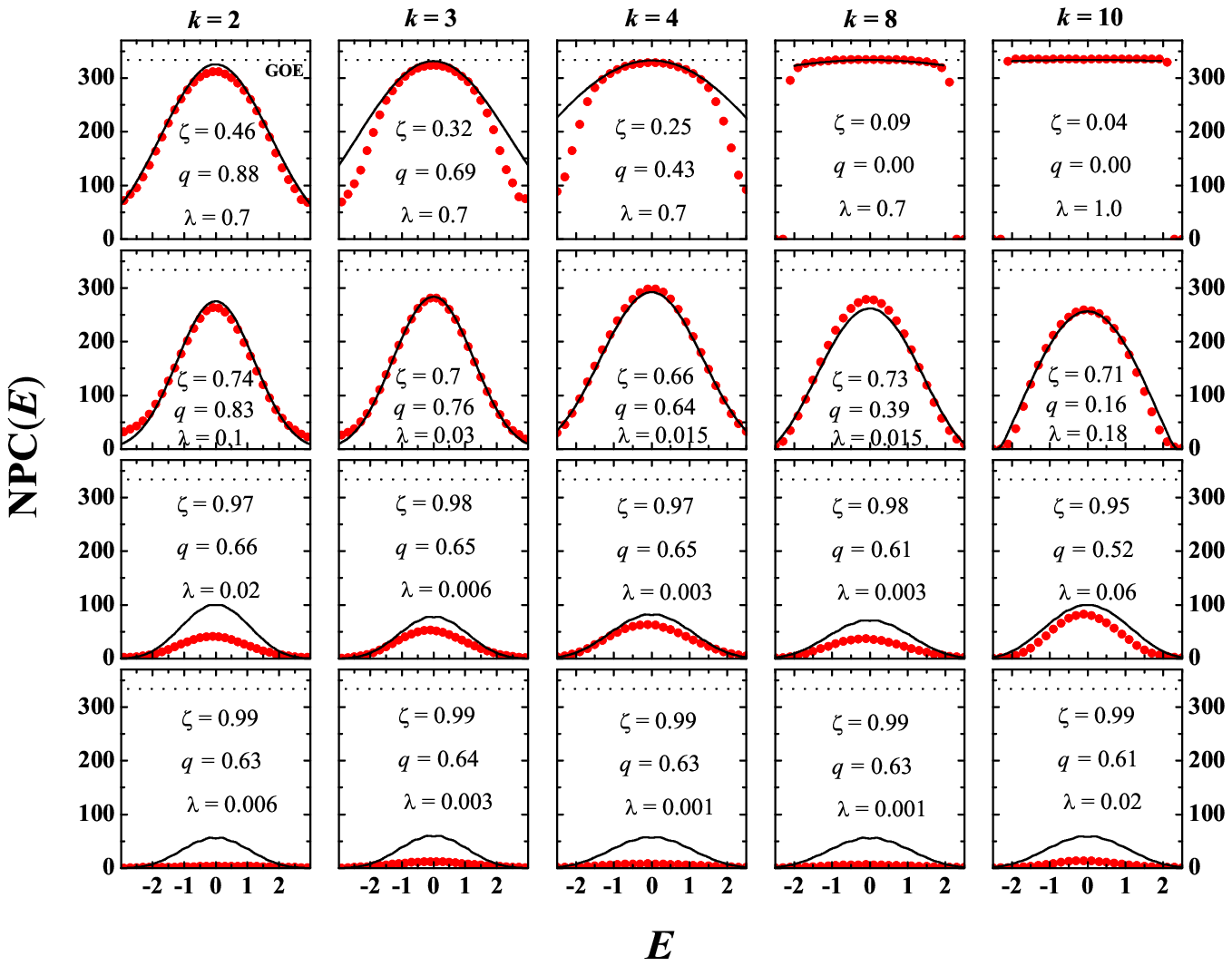}
\caption{Ensemble averaged NPC as a function of normalized energy $E$ for a 100 member BEGOE(1+$k$) with $m=10$ interacting bosons in $N=5$ sp states for different values of $k$. Ensemble averaged BEGOE(1+$k$) results are represented by solid circles while continuous curves
correspond to the theoretical estimates in the chaotic domain obtained using Eq.~\eqref{eq.npc}. The ensemble averaged $\zeta$ and $q$ values are also given in the figure. GOE estimate is represented by dotted line in each graph.}
 	\label{fig-npc}	
 \end{figure}	

\begin{figure}[tbh!]
\centering
\includegraphics[width=0.75\textwidth]{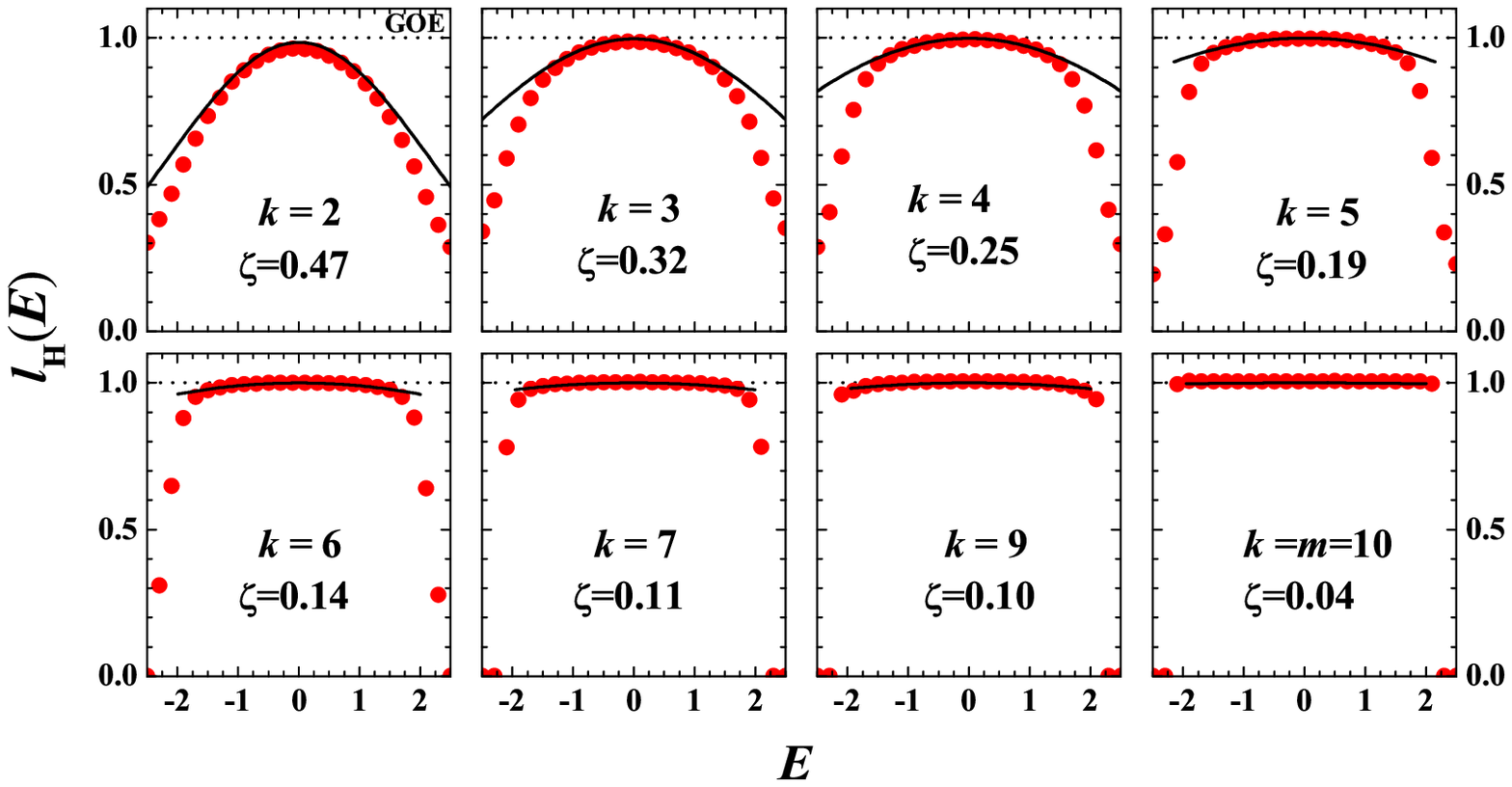}
\caption{Ensemble averaged localization lengths $l_H$ vs. normalized energy $E$ for a 100 member BEGOE(1+$k$) with $m=10$ interacting bosons in $N=5$ sp states for different $k$ values. Here, $\lambda=1$ is chosen for all $k$. Ensemble averaged BEGOE(1+$k$) results (red solid circles) are compared with the smooth forms obtained via Eq.\eqref{eq.lh} involving parameters $\zeta$ and $q$. The ensemble averaged $\zeta$ values are given in the figure and Eq.\eqref{eq.ent9} is used for $q$ values. Dotted lines in each graph represent GOE estimate.}
 	\label{fig-lh} 	
 \end{figure}

\section{Conclusions}
\label{sec:7}

In the present work, we have analyzed wavefunction structure of dense many-body bosonic systems with $k$-body interaction by modeling the Hamiltonian of these complex systems using BEGOE(1+$k$). We have shown that for dense boson systems with BEGOE(1+$k$), the $q$-polynomials are used to describe the transition from Gaussian to semi-circle in the state density as the strength of the $k$-body interaction increases. A complete analytical description of the correlation coefficient $\zeta$, which is related to variance of strength functions, is obtained in terms of $N$,$m$,$k$ and $\lambda$ and it is found to describe the embedded ensemble results very well for all the values of rank of interaction $k$. Also, in the dense limit $\zeta \rightarrow 0$. We have also obtained formula for $\lambda_t$ in terms of ($m$, $N$, $k$). Further, it is shown that in the strong interaction domain ($\lambda >> \lambda_t$), the strength functions make transition from Gaussian to semi-circle as the rank of interaction $k$ increases in BEGOE(1+$k$) and their smooth forms  can be represented by the $q$-normal distribution function $f_{CqN}$ to describe this crossover. Moreover, the variation of the lowest four moments of strength functions computed numerically are in good agreement with the analytical formulas obtained in \cite{KM2020c}. With this, we have first utilized the interpolating form for strength function $f_{CqN}$ to describe the fidelity decay in dense boson systems after $k$-body random interaction quench. Secondly, using smooth forms for $f_{qN}$ and $f_{CqN}$, we have also derived two parameter ($q$ and $\zeta$) formula for NPC valid in thermalization region and shown that these smooth forms describe BEGOE(1+$k$) ensemble averaged results very well. Therefore, the results of this work, along with \cite{Manan-Ko,KM2020,KM2020c}, establish that the $q$-Hermite polynomials play a very significant role in analyzing many-body quantum systems interacting via $k$-body interaction. The generic features explored in this work are important for a complete description of many-body quantum systems interacting via $k$-body interaction as the nuclear interactions are now known to have some small 3-body and 4-body parts  and higher body interactions may become prominent in strongly interacting quantum systems \cite{Cotler-2017,Blatt,Hammer}.

Following the work in \cite{Lea-power}, it is interesting to analyze power-law behavior of fidelity decay for very long time using embedded ensembles with $k$-body forces as smooth forms of strength functions can be represented by $f_{CqN}$. Further, as smooth forms for the density of states can be represented by $f_{qN}$, it is possible to study normal mode decomposition of the density of states for various $k$ values using $f_{qN}$ \cite{Brody,MF,leclair} and thereby one can study spectral statistics in strongly interacting quantum systems. This is for future. It is also known that the strength functions and  the entanglement essentially capture the same information about eigenvector structure \cite{CMejia1998,brown2008} and therefore it is important to study entanglement properties using embedded ensembles with $k$-body forces. 
 
\section*{Acknowledgements}
Thanks are due to Manan Vyas for collaboration in the initial stages of this work and V. K. B. Kota for many useful discussions. Authors acknowledge support from Department of Science and Technology(DST), Government of India [Project No.: EMR/2016/001327]. NDC acknowledges support from the International Centre for Theoretical Sciences (ICTS) during a visit for participating in the program -  Thermalization, Many body localization and Hydrodynamics (Code: ICTS/hydrodynamics2019/11).

\end{document}